\documentstyle[prl,aps,multicol,floats]{revtex} 

\input epsf

\begin{document}


\title{Freezing by Monte Carlo Phase-Switch}

\author{N.B. Wilding and A.D. Bruce} 
\address{Department of Physics and Astronomy, The University of
Edinburgh\\
Edinburgh, EH9 3JZ, Scotland, United Kingdom} 
\maketitle

\makeatletter
\global\@specialpagefalse
\def\@oddhead{}
\let\@evenhead\@oddhead
\def\@oddfoot{\reset@font\rm\hfill \thepage\hfill
} \let\@evenfoot\@oddfoot
\makeatother

\begin{abstract}

We describe a Monte Carlo procedure which allows sampling of the
disjoint configuration spaces associated with crystalline and fluid
phases, within a single simulation.  The method utilises biased
sampling techniques to enhance the probabilities of gateway states (in
each phase) which are such that a global switch (to the other phase)
can be implemented. Equilibrium freezing-point parameters can be
determined directly; statistical uncertainties prescribed
transparently; and finite-size effects quantified systematically.  The
method is potentially quite general; we apply it to the freezing of
hard spheres.


\end{abstract}

\newpage
\twocolumn

Freezing is the archetypal phase transition, one of the prime
examples of thermodynamics in action, and a topic of ongoing interest 
\cite{kegel,schmidt,kofke}.
It is therefore remarkable that the challenge it presents to
computational science has yet to be satisfactorily met. The generic
problem is to compute the location of the freezing transition (more
generally the liquid-solid coexistence curve) on the basis of a
particle-level model. The approach to this problem has
evolved little since the pioneering work of Hoover and Ree
\cite{hooverree}. The free energy of each phase (fluid (F)
and crystalline solid (CS)) is computed for states of a range of
densities, using integration methods which connect their thermodynamic
properties with those of effectively single-particle reference states,
whose free energies are known {\it a priori}; the two branches of the
free-energy are then matched to
determine the freezing parameters. This approach has several
drawbacks. The integration path may encounter singularities --both
real and artificial \cite{singularities}.  Corrections may be needed
to allow for the fact that the path does not quite reach the idealised
reference state \cite{frenkelladd}.  The implicit perspective adopted
(that there are two separate calculations to be done --one for each phase)
has meant that predictions for freezing parameters are often a
synthesis of work done by different authors on different system-sizes,
making it hard to quantify finite-size effects
\cite{fseffects,frenkelsmit}.

This paper describes a different approach to the problem.  We build
on recent work \cite{LSpapers}, in which we showed that the
disjoint configuration spaces associated with two phases of a
many-body system can {\em both} be visited in a {\em single} Monte Carlo
(MC) simulation, by harnessing
extended-sampling (ES) methods \cite{ESmethods} to facilitate a direct
switch from one phase to the other: instead of {\em traversing} a
region where both phases coexist \cite{doingthetraverse} the method
may be thought of as {\em leaping} from one space to the other; the
role of ES is to allow the system to find the `gateway' states from
which a leap will be accepted. The method was developed
\cite{LSpapers} to tackle the problem posed by two crystalline phases,
where interfacial states are computationally problematic. The same is
true of the CS-F problem. But significant extensions of the framework 
are needed to address the qualitatively different characters of the
two configuration spaces. 
First, the {\em communal entropy} of the fluid
\cite{communalentropy} provides a conceptually different  form of
barrier that has to be negotiated to reach the gateway states: we show
how one can do this. Second, the {\em distinct contributing
configurations} have to be identified with care: in so doing we
unearth a small but significant flaw, inherent (we think) in all
previous simulation-studies of CS-phase free energies. The method we
develop is general;  we illustrate it here with a study of the
entropically-driven freezing of hard spheres, where  earlier
studies provide useful benchmarks 
\cite{hooverree,frenkelsmit,speedy}

Consider $N$ particles (hard spheres) confined to  volume $V$, 
variable under a constant external effective pressure $p$ 
\cite{units}, and subject to periodic boundary conditions.
The configurational weight of a phase may be written as
\begin{equation}
\label{eq:zcaldef}
{\cal Z}_{\gamma}(N,p)  = \int_{0}^{\infty} dV e^{-p V} Z_{\gamma}(N,V)
\end{equation}
where $\gamma$ (CS or F) labels the phase, while
\begin{equation}
\label{eq:zsimpdef}
Z_{\gamma}(N,V)
= C_{0} \prod_{i=1}^{N}\int_{V,\gamma} d \vec{r}_{i} 
e^{-E( \{\vec{r}\})} 
\end{equation}

Here $E$ is the hard sphere configurational energy \cite{units}.
The prefactor $C_{0}$ is chosen according to whether the particles are
regarded as `strictly classical' ($C_0=1$) or `classical but
indistinguishable' ($C_0=\frac{1}{N!}$). The results for 
observables are independent of this choice.  
The $\gamma$-label on the integral stands for some {\em
configurational constraint} that picks out configurations $\{\vec{r}\}$ that
`belong' to phase $\gamma$. We choose to formulate that constraint as
follows \cite{constraintissues}. Let $\vec{R}_{1}^{\gamma}
\ldots \vec{R}_{N}^{\gamma} \equiv \{ \vec{R}\}^{\gamma}$ denote some 
{\em representative configuration} of phase $\gamma$. Then the constraint may
be regarded 
as picking out those configurations which can be
reached from $\{ \vec{R}\}^{\gamma}$ on the simulation timescale \cite{moreontimescales}.
It is convenient to use the sites defined by $\{ \vec{R}\}^{\gamma}$ as the 
origins 
of the 
particle coordinates. Thus we define a set 
of displacement vectors $\{\vec{u}\}$ by
$\vec{u}_i\equiv \vec{r_i} - \vec{R}^{\gamma}_i$
and write $E^{\gamma} (\{\vec{u}\} ) \equiv E (\{ \vec{R}^{\gamma} + \vec{u}\} )$.

In the case of the F-phase {\em all} 
contributing configurations are reachable from any one; we 
may write simply
\begin{equation} 
\label{eq:zsimpF}
Z_{F}(N,V)
= C_{0} \prod_{i=1}^{N}\int_{V,\{ \vec{R}\}^{F}} d \vec{u}_{i} e^{-E ^F (\{\vec{u}\})} 
\end{equation}
where $\{ \vec{R}\}^{F}$ is some specific but arbitrary fluid configuration, 
which can be  selected at random in the course of MC exploration of the 
fluid phase. It is natural to choose 
$\{ \vec{R}\}^{CS}$ to 
define the sites of a lattice of the appropriate symmetry (here {\it fcc})
and scale \cite{andscale}. 
But one must recognise that the complete CS configuration space
actually comprises a number of  
distinct mutually inaccessible {\em fragments} 
\cite{theyaredistinct} 
corresponding essentially 
to the different permutations 
of particles amongst lattice sites \cite{alexander}.
By symmetry each fragment  should contribute equally to the configurational
weight; but MC simulation will visit (and thus count) only the states within 
the fragment in which it is initiated. The total 
configurational weight of the CS phase is given by multiplying the
contribution of one fragment by the number of fragments. 
Since global translation
(permitted by the boundary conditions) ensures that {\em one} fragment 
includes {\em all} possible locations of any chosen particle,
the number of fragments is the number of ways of 
assigning the `other' $N-1$ particles to $N-1$ Wigner-Seitz cells
of some underlying notional fixed lattice.
This number is not $N!$ but $(N-1)!$. Thus
\begin{equation}
\label{eq:zsimpCS}
Z_{CS}(N,V)
= C_{0} (N-1)!\prod_{i=1}^{N}\int_{V,\{ \vec{R}\}^{CS}} d \vec{u}_{i} 
e^{-E^{CS}(\{\vec{u}\})} 
\end{equation}
The ratio of the configurational weights of the two phases (the ratio of their total {\em a priori} probabilities) follows by
combining Eqs.~\ref{eq:zcaldef},~\ref{eq:zsimpF} and \ref{eq:zsimpCS}:
\[
{\mathcal{R}}_{\mbox{\sc {f,cs}}}
=\frac{P(F| N, p)}{P(CS|N,p)}
=\frac
{{\cal Z}_{F}(N,p)}  
{{\cal Z}_{CS}(N,p)}
\]
\begin{equation}
\label{eq:zrat}
=\frac
{[N!]^{-1} \int_{0}^{\infty} dV e^{-p V} 
\prod_{i=1}^{N}\int_{V,\{ \vec{R}\}^{F}} d \vec{u}_{i} e^{-E^{F}(\{\vec{u}\})} }
{[N]^{-1}  \int_{0}^{\infty} dV e^{-p V} 
\prod_{i=1}^{N}\int_{V,\{ \vec{R}\}^{CS}} d \vec{u}_{i} e^{-E^{CS}(\{\vec{u}\})} }
\end{equation}
from which the Gibbs free-energy-density difference follows as 
\begin{equation}
\label{eq:gibbs}
\Delta g \equiv g_{CS}(N,p) -g_{F} (N,p)  \equiv \frac{1}{N} \ln {\mathcal{R}}_{\mbox{\sc {f,cs}}}
\end{equation}

In writing Eq.~\ref{eq:zrat} we have chosen to split the fragment
number $(N-1)!$ into separate factors of $1/N!$ and $1/N$. If one
so wishes \cite{matterofchoice} one may regard the former as the
familiar indistinguishability overcount-correction appropriate for phases
(fluids) of non-localised particles.  But then one must recognise the
existence of an  analogous correction (the $1/N$) for the CS phase,
in which particles are localised ---but only relative to one another.
It seems that this correction has been missed by other authors; 
we shall see that it contributes significantly to finite-size effects.

The relative stability of the two phases is determined by the ratio of
the associated configurational weights (Eq.~\ref{eq:zrat}).  To
determine that ratio we need a MC procedure which visits both solid
and fluid regions of configuration space. Since, by construction, the
system may be transformed between the CS and F reference states simply
by switching the representative vectors ($\vec{R}_i^F
\rightleftharpoons \vec{R}_i^{CS}\, \forall i$), 
by continuity, {\em any} CS (F) configuration `sufficiently close' to  
the representative one will also transform to a F (CS) state under this 
operation. This phase  switch can itself be realised as a MC step, so that
the phase label $\gamma$ becomes a stochastic variable.
The set of configurations for which the MC switch will be {\em accepted}
will, however, constitute only a small 
fraction of the respective configuration spaces.
To ensure effective two-phase sampling the MC procedure must be 
biased \cite{ESmethods} to enhance the probabilities with which these 
`gateway' regions are visited. To that end we define an order parameter
\[
M=M_{\gamma}(\{\vec{u}\}) = \sum_i 
\left\{
O_{i} [1-\theta(u_i -u_c)]
+ 
T_{i} \theta(u_i -u_c)
\right\}
\]
Here $\theta$ is the step function.  
$T_{i}\equiv \alpha u_i$ measures the length of a notional 
tether connecting site $i$ to its associated particle 
\cite{historyoftethers}.
$O_{i}$ measures the overlap (between particle $i$ and its
neighbours) which {\em would} be created by a phase switch.  The
parameter $\alpha$ controls the relative importance of $T_i$ and $O_i$; 
$u_c$ controls the tether-length domain in which
each contributes \cite{valuesofparameters}.
The equilibrium states of both phases are characterised 
by large $M$ values. The `overlap' term contributes in both phases:
swapping the ${\{ \vec{R}\}}$ vectors will (in general) produce a configuration
of the `other' phase in which spheres overlap.
The `tether' term contributes only in the F-phase 
\cite{givenconventions} where particles may drift arbitrarily 
far from the sites with which they are nominally associated;
the tethers  provide the means to `pull' the fluid up the communal 
entropy barrier.
We identify the gateway states as those which have $M=0$ (ie $O_i=0$ and $u_i 
< u_c$, $\forall i$).
The constraint that $M=0$ imposes on the overlap  
simply recognises that MC-switches which generate overlaps
will necessarily be rejected.
The constraint ($u_i < u_c$) on the tether length 
is needed
to ensure that switches from the fluid create only
{\em crystalline} solid (not defective, glassy) configurations.
The entire region of configuration  space relevant to the problem 
can then be sampled in the multicanonical ensemble defined by
\begin{equation}
\label{eq:zmultican}
\tilde{\cal{Z}}(N, p, {\{\eta\}} \equiv
\sum _{\gamma} \int_{0}^{\infty} dV 
\prod_{i}^{N}\int_{\gamma} d \vec{u}_{i} 
e^{-{\cal H}^{\gamma}( \{\vec{u}\} , V)}
\end{equation}
where
\[
{{\cal H}}^{\gamma}( \{\vec{u}\} ,  V) = E^{\gamma}( \{\vec{u}\} ) + p V +\eta_{\gamma} (M) -
\delta_{\gamma,CS}\ln{(N-1)!}
\]
while $\{\eta\}$ represents weights (defined on the M-macrostates)
which have to be constructed so as to enhance, appropriately, the
probabilities of the $M=0$ gateway states \cite{multicanweights}.
Simulation in this ensemble allows one to {\em measure} the
multicanonical probability distribution $P(M, V, \gamma |N, p, \{\eta\} )$ from which
(unfolding the bias due to the weights) one may {\em infer} the
true equilibrium distribution $P(M, V, \gamma |N, p)$.  The desired ratio of the
phase probabilities (Eq.~\ref{eq:zrat}) follows by `marginalising'
$M$ and $V$ to give the {\em a priori} probabilities of the phases. Having
the underlying distribution of $M$ {\em and} $V$ allows one to
determine, in addition, the value of ${\mathcal{R}}_{\mbox{\sc {f,cs}}}$ at neighbouring
pressures, using histogram reweighting techniques
\cite{histogramreweighting}. 

We turn to the MC procedure required for efficient exploration of the
space spanned by the configuration variables $\{\vec{u}\}$, $V$ and
$\gamma$. It comprises four types of configuration update, each of
which is accepted with a probability defined by a Metropolis rule
\cite{frenkelsmit} and reflects the associated change in the effective 
energy ${\cal H}$. The first two --particle position updates
\cite{keeponefixed} and volume updates (implemented as dilations)-- 
are effected in standard ways \cite{frenkelsmit}. The 
third --like the first two-- also preserves the phase label; but it is novel. 
In this process, we choose two sites at random ($i$ and $j$ say) and
identify the corresponding displacement vectors $\vec{u}_i$ and
$\vec{u}_j$.  The candidate configuration is defined by the
replacements
\[
\vec{u}_i \rightarrow \vec{u}_i^{\prime} \equiv\vec{u}_j + \vec{R}_j - \vec{R}_i 
\hspace*{0.5cm}
\mbox{and}
\hspace*{0.5cm}
\vec{u}_j \rightarrow \vec{u}_j^{\prime} \equiv \vec{u}_i + \vec{R}_i - \vec{R}_j
\]

This process can be thought of as an {\em association} update: the
particle initially associated with (`tethered to') site $j$ is subsequently associated
with site $i$ (and {\it vice versa}).  It changes the {\em
representation} of the configuration (the coordinates $\{\vec{u}\}$); but it
leaves the physical configuration invariant.  It
is {\em required} in the fluid phase only \cite{candoitinCS}. In the
fluid phase the particles diffuse far from the sites with which they
are initially associated; the members of $\{\vec{u}\}$ become large and the
tethers correspondingly so; association updates allow the tethers to
respond efficiently to the influence of the tether contribution to
$\{\eta\}$.  Finally, the `phase update' (the switch) entails
replacing one set of representative vectors, $\{ \vec{R}\}^{\gamma}$ say, by the
other, $\{ \vec{R}\}^{{\gamma}^{\prime}}$, with the volumes scaled appropriately and
the displacement coordinates $\{\vec{u}\}$ held fixed \cite{volumecalingfraction}.

Simulations have been performed using systems of $N=32$, $108$ and
$256$ particles.  Figure~\ref{fig:rhodist} shows the density
distribution for the $N=256$ system in the vicinity of the coexistence 
pressure.  Coexistence ($\Delta g=0$; Eq.~\ref{eq:gibbs}) is identified
by the equality of the
contributions associated with each phase (essentially the area under
each peak).

Figure~\ref{fig:pvalue} shows the coexistence pressure for our three
system sizes plotted as a function of $1/N$ \cite{whythis}.  The
results for $N=108$ and $N=256$ were obtained in the fashion just
described; the associated uncertainties $\sigma [p]$
follow simply from Eq.~\ref{eq:gibbs} as 
$\sigma [p] = \sigma [{\cal R}]/(N\mid \Delta v \mid )$ where $\Delta v =[\bar{V}_F-\bar{V}_{CS}]/N$ and $\sigma [ {\cal R} ]$ 
is the uncertainty in the measured ratio of the peak-weights, which
is controlled, at heart, by the statistics
of the inter-phase switch. The result for $N=32$ was determined
differently: this system is sufficiently small that  
transitions back and forth between
F and CS phases occur {\em spontaneously}, over a range of pressures,
and a density distribution (sampling both phases) can be determined
--and a coexistence pressure inferred --{\em without} multicanonical
weighting. The three points are consistent with the presumed scaling form
\cite{whythis}. The extrapolated prediction
($p =11.49(9)$) is, within error, in accord with \cite{hooverree}
and \cite{speedy} (see Fig.~\ref{fig:pvalue} inset).

The lower set of data points shown in figure~\ref{fig:pvalue} gives
the values of the coexistence pressure implied by our measurements for
$N=108$ and $N=256$ if one fails to fold in the $1/N$ correction in
Eq.~\ref{eq:zrat}. The associated {\em overestimate} of the
CS-configurational weight {\em lowers} the predicted coexistence
pressure by an amount ($[\ln N]/[N \Delta v]$) 
which
is significant for systems of this
size, and leads to values which it is hard to reconcile 
(cf the dashed line in Fig.~\ref{fig:pvalue}) with the
independent measurement at $N=32$ \cite{wrongwayround}.
While this correction vanishes in the $N\rightarrow \infty$ limit, its 
existence is potentially important for any systematic study of 
the finite-size scaling of free energies \cite{polson}.

We summarise. We have presented a method which allows one to locate
liquid-solid coexistence parameters (and uncertainties) directly and
transparently (Fig.~\ref{fig:rhodist}) within a {\em single}
simulation, conducted in the {\em appropriate} (constant pressure)
ensemble. The method avoids the need to appeal to integration through
to `distant' reference states, double-tangent-constructions or off-the-shelf
equations of state.  It prescribes finite-size effects explicitly and
handles systems sufficiently large (cf \cite{kegel}) that the limiting
thermodynamic behaviour can be identified with some confidence.  The method can
be readily generalised to systems with real (soft) potentials
\cite{kofke} and arbitrary geometries \cite{schmidt}.  It can also be
naturally combined with histogram reweighting techniques
\cite{histogramreweighting} 
to allow the full coexistence-curve to be mapped efficiently.

 \newpage
\newcounter{abc}
\renewcommand{\thefigure}{\arabic{figure}\alph{abc}}

\begin{figure}[t] 
\leavevmode
\epsfxsize=70mm 
\begin{center}
\epsffile{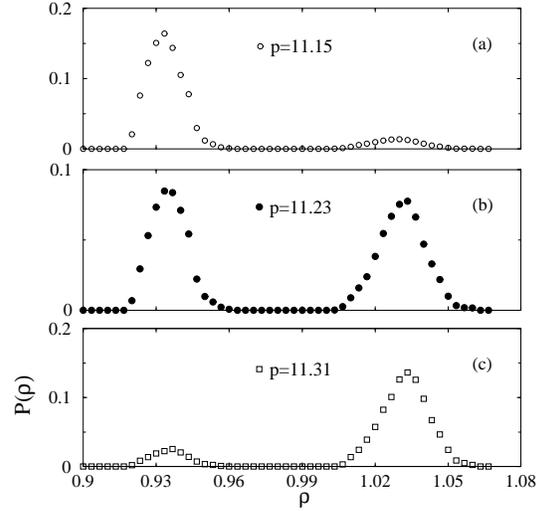}
\end{center}
\caption
{The distribution of the density of the system of $N=256$ particles at 
pressures (a) just below, (b) at and  (c) just above coexistence for this $N$.
The mean single phase density averages are $\rho_F=0.934(3)$ and 
$\rho_{CS}=1.031(4)$ in accord with the coexistence 
parameters reported in \protect\cite{speedy}.
}
\label{fig:rhodist}
\end{figure}

\begin{figure}[t]
\leavevmode
\epsfxsize=70mm 
\begin{center}
\epsffile{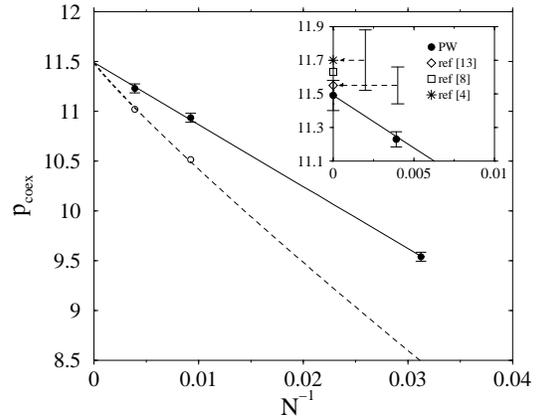}
\end{center}
\caption
{The coexistence pressure for systems of different $N$ using 
Eq.~\ref{eq:zrat} both with ($\bullet$) and without ($\circ$) 
the $1/N$ prefactor in the CS configurational weight. 
The solid lines is a fit to the former; the dashed line is lower by
$\ln N/ [N\Delta v]$. The inset compares our extrapolated 
value with the results of others, with error bars shifted for clarity.}
\label{fig:pvalue}
\end{figure}




\begin{references}
\bibitem{units} 
$p$, $V$ and $E$ are measured in units of $kT/D^3$ and $D^3$
and $kT$ respectively, where $D$ is the hard sphere diameter.
\bibitem{moreontimescales} The timescale is presumed to be sufficiently 
long to explore `one phase' but still short compared to (unaided)
interphase traverses.
\bibitem{theyaredistinct} At our starting point (Eq.~\ref{eq:zsimpdef})
the indices $1 \ldots N$  on the coordinates $\protect{\{\vec{r}\}}$ are (can only be)
{\em particle} labels which confer distinguishability. Any `correction for
indistinguishability' is made {\em post hoc} through $C_0$. 
With this perspective (there are alternatives), 
the `fragments' are `distinct'.
\bibitem{valuesofparameters} There is some licence
in setting the values of these parameters; we chose
$u_c=D$ and $\alpha = 1.7$.
\bibitem{volumecalingfraction} In the switch,  the volume is scaled by 
$\alpha_v \equiv \bar{V}^{{\gamma}^{\prime}}/\bar{V}^{\gamma}$   
where $\bar{V}_{\gamma}$ is the equilibrium volume of phase $\gamma$.
\bibitem{whythis} This choice of scaling variable is common 
--see eg \cite{polson}-- but seems to lack
a rigorous basis.
\bibitem{wrongwayround} 
The inset of \protect~\ref{fig:pvalue} shows that our extrapolation
lies below the results of \cite{hooverree,frenkelsmit,speedy}.  If
there is a trend here the $1/N$ correction is not the source: it has
the opposite effect. The diiferences may lie in residual finite size 
effects in  equations of state fitted to fluid simulation data.
\end{references}
\end{document}